\documentclass[twocolumn]{esrel2021}

\def\papername{\jobname}

\usepackage{lipsum} 
\usepackage{booktabs}
\let\captionbox\undefined
\usepackage[font=small,labelfont=bf]{caption}
\usepackage{multirow}
\usepackage{color, colortbl}
\definecolor{Gray}{gray}{0.9}
\usepackage[authoryear]{natbib}
\begin{document}

\markboth{L.A. Jimenez-Roa, T. Heskes, and M. Stoelinga}{Fault Trees, Decision Trees, And Binary Decision Diagrams: A Systematic Comparison}

%%%%%%%%%%%%%%%%%%%%%%%%% Plase keep this command for single column for abstract section.
\twocolumn[
%%%%%%%%%%%%%%%%%%%%%%%%%

\title{Fault Trees, Decision Trees, And Binary Decision Diagrams: A Systematic Comparison}

\author{Lisandro A. Jimenez-Roa}
\address{Formal Methods and Tools (FMT), University of Twente, The Netherlands. \email{l.jimenezroa@utwente.nl}}

\author{Tom Heskes}
\address{Institute for Computing and Information Sciences, Radboud University Nijmegen, The Netherlands. \email{Tom.Heskes@ru.nl}}

\author{Marielle Stoelinga}
\address{Formal Methods and Tools (FMT), University of Twente, The Netherlands. \email{m.i.a.stoelinga@utwente.nl}}

\begin{abstract} 
In reliability engineering, we need to understand system dependencies, cause-effect relations, identify critical components, and analyze how they trigger failures. Three prominent graph models commonly used for these purposes are fault trees (FTs), decision trees (DTs), and binary decision diagrams (BDDs).
These models are popular because they are easy to interpret, serve as a communication tool between stakeholders of various backgrounds, and support decision-making processes. Moreover, these models help to understand real-world problems by computing reliability metrics, minimum cut sets, logic rules, and displaying dependencies.
Nevertheless, it is unclear how these graph models compare. Thus, the goal of this paper is to understand the similarities and differences through a systematic comparison based on their (i) purpose and application, (ii) structural representation, (iii) analysis methods, (iv) construction, and (v) benefits \& limitations. Furthermore, we use a running example based on a Container Seal Design to showcase the models in practice.
Our results show that, given that FTs, DTs and BDDs have different purposes and application domains, they adopt different structural representations and analysis methodologies that entail a variety of benefits and limitations, the latter can be addressed via conversion methods or extensions. Specific remarks are that BDDs can be considered as a compact representation of binary DTs, since the former allows sub-node sharing, which makes BDDs more efficient at representing logical rules than binary DTs. It is possible to obtain cut sets from BDDs and DTs and construct a FT using the (con/dis)junctive normal form, although this may result in a sub-optimal FT structure.

%However, Binary DTs and not Ordered BDDs are pretty simillar. Moreover, unlike DTs, FTs and BDDs model causality. DTs and BDDs share a similar type of elements, and information propagation. 
\end{abstract}


%%%%%%%%%%%%%%%%%%%%%%%%% Please keep this closing bracket to complete the single column format for abstract.
]
%%%%%%%%%%%%%%%%%%%%%%%%%
% -------------------- %
\section{Introduction}
Three commonly used graph models are Fault Trees (FTs), Decision Trees (DTs), and Binary Decision Diagrams (BDDs). These models are popular because they provide a graphic representation of a hierarchical data structure. They are widely used in different domains and applications such as reliability engineering, system analysis, and computer memory optimization \citep{lee1985fault,rokach2008data,kubica2021binary} (see Table \ref{tb:comparision_between_models}).

This paper aims at understanding the explicit similarities and differences between these models. This comparison is important, for example, to facilitate the selection of a model for a given application, or to clarify terminology such as ``Tree'' or ``Decision'' for less experienced users. To this end, we focus on comparing these models systematically, as well as discuss conversion methods to transform one model into another.

The structure of this paper is as follows. Section \ref{sec:methodology} describes our methodology, and Sections \ref{sec:FT} to \ref{sec:BDD} present the results per model: FTs, DTs and BDDs, respectively. In Section \ref{sec:conversion_methods}, we discuss the conversion methods, and finally, Sections \ref{sec:discussion} and \ref{sec:conclusions} contain the discussion and conclusions.

% Comparison table 
\begin{table*}[!h]
\centering
    \caption{Comparison between FTs, DTs and BDDs.}
    \label{tb:comparision_between_models}
    \includegraphics[width=1\linewidth]{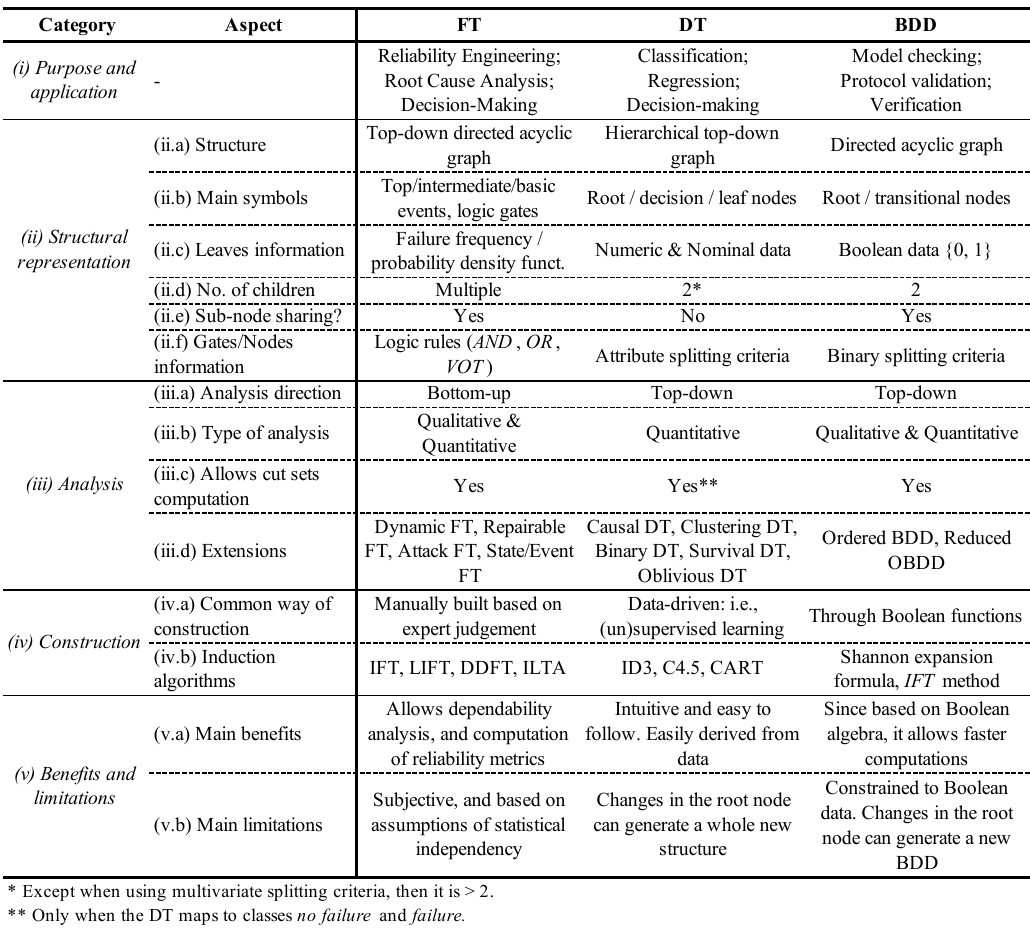}
\end{table*}
% -------------------- %
\section{Methodology}\label{sec:methodology}
Our comparison follows a systematic approach proposed by \cite{smith2017understanding}. Our main steps are (1) to propose five main categories encompassing different aspects of interest, namely (i) purpose \& application, (ii) structural representation, (iii) analysis, (iv) construction, and (v) benefits \& limitations; (2) to refine each category into aspects; (3) to adapt a running example from \cite{stamatelatos2002fault} consisting of a FT for a \textit{Container Seal Design}, and model it by DTs and BDDs; (4) to discuss conversion methods to transform one model into another (here it is important to clarify that minimum compatibility requirements must be met between models, e.g., DTs must model diagnostic decision rules and binary variables); and (5) summarize the different aspects per model.

% -------------------- %
\section{Fault Trees}\label{sec:FT}

\subsection{Purpose and application}

Fault Tree Analysis (FTA) is a key method in reliability engineering and root cause analysis, to support decisions in system design and maintenance. FTA is ISO standardized \citep{international2006iec} and has been used in a wide range of domains including automotive, aerospace, and nuclear industries \citep{kabir2017overview}.

Technically, a Fault Tree (FT) is a directed acyclic graph that models why a system fails, by identifying how low-level failures propagate through the system and lead to the system-level failure.

% -----------------------------------
% Structural definition of a FT:
\subsection{Structural representation}

\begin{figure}[!h]
\centering
\includegraphics[width=1\linewidth]{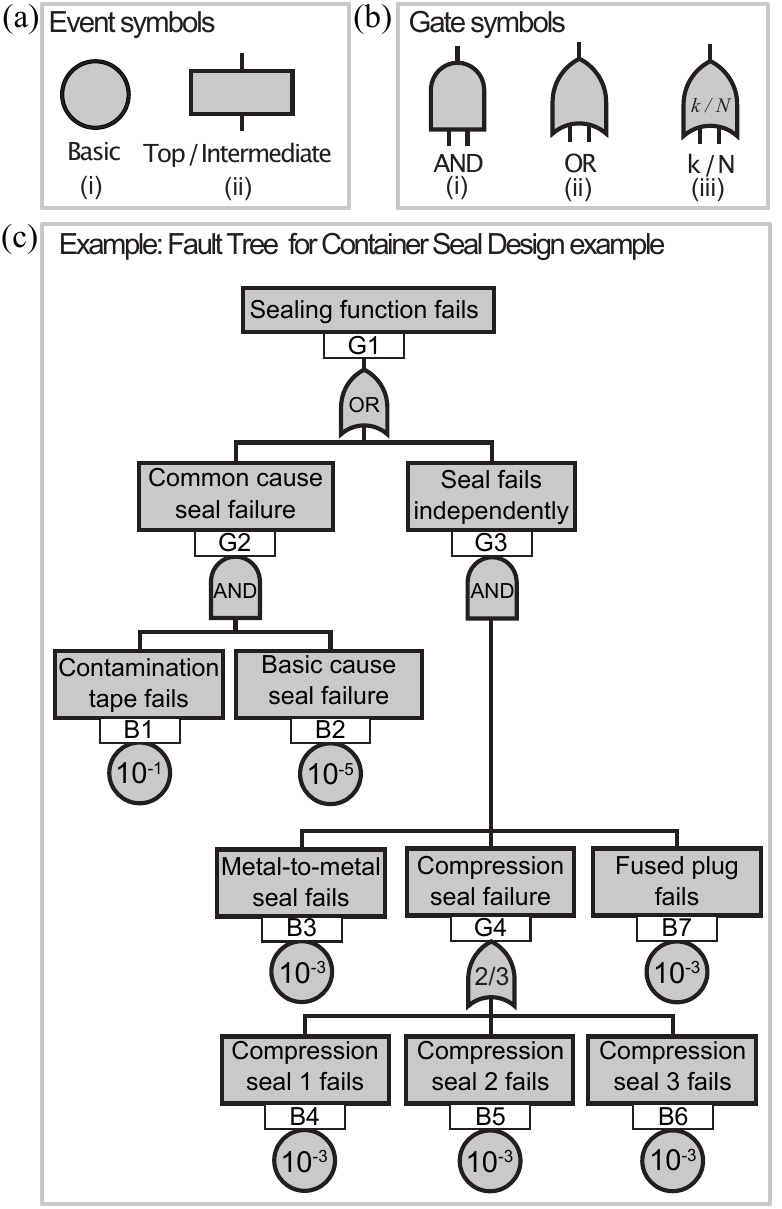}
\caption{Elements in an FT: (a) event symbols, (b) gate symbols, and (c) example of an FT, adapted from: \cite{stamatelatos2002fault}.}
\label{fig:static_fault_tree_structure}
\end{figure}

FTs are composed of different symbols (Fig. \ref{fig:static_fault_tree_structure}(a-b)), whose objective is to model the logic relations between the events (basic and intermediate, Fig. \ref{fig:static_fault_tree_structure}.(a)) and the top event. Logical transitions between events must meet the conditions set by the \textit{gate symbols} (Fig. \ref{fig:static_fault_tree_structure}.(b)).\\

% Figure where is summarized the main components in the architecture of a FT

FTs distinguish several event types (Fig. \ref{fig:static_fault_tree_structure}.(a)):

\begin{romanlist}[(iii)]

    \item \textit{Basic events}, symbolized by a circle, corresponds to an event that initiates the failure of the system in question. The basic events do not require further refinement; in other words, the resolution achieved is adequate.
    
    \item \textit{The top event}, symbolized by a rectangle, is localized at the top of the FT. It models the failure of the complete system under consideration. \textit{Intermediate events} are associated with (sub)system/component failures, and are equipped with a logical gate. 

    %The \textit{Top event} is always at the top of the FT diagram and are associated with the undesired ruling event. The Top event occurs when its associated logic gate is activated.
    
    %\item \textit{Conditioning event}: symbolized by an oval, and it is used to specify a condition or restriction that needs to be fulfilled to activate a gate. 
    
    %\item \textit{Undeveloped event}: symbolized by a diamond, and corresponds to those events that cannot be further refined either due to insufficient knowledge/data, or because the refinement is meaningless.
    
    %\item \textit{External event}: symbolized by a ``house'' symbol, and associated to events that are expected to occur, therefore it is not considered a fault itself.
    
\end{romanlist}

Three commonly used gate types in FTs are (Fig. \ref{fig:static_fault_tree_structure}.(b)):

\begin{romanlist}

    \item \textit{AND gates} indicate that the associated event will only occur if \textit{all} the associated (basic or intermediate) events occur.
    
    \item \textit{OR gates} indicate that the associated event will only occur if \textit{at least} one of the associated (basic or intermediate) events occur.
    
    \item \textit{k/N gates} (or \textit{VOT gates})  indicate that the associated event will occur minimally $k$ of the $N$ the events associated to the gate will occur.
    
    %\item \textit{INHIBIT gate}: symbolized by a hexagon. It is a special case of the AND gate, where a condition must also be met before the input can produce an output.
    
    %\item \textit{PRIORITY AND (PAND) gate} (Figure \ref{fig:static_fault_tree_structure}.b.V): symbolized by a shield with a flat base with a circumscribed triangle. This gate is activated with the \textit{condition} that all the associated events occur in a specific sequence.
    
    %\item \textit{EXCLUSIVE OR (XOR) gate}: symbolized by a shield with a curved base with a circumscribed triangle. This gate is activated only if exactly one of the input events occurs. The difference to the usual or inclusive OR gate (Figure \ref{fig:static_fault_tree_structure}.b.II) in the situation where both input events occur is excluded.
    
\end{romanlist}

%\textit{Transfer symbols} (Fig. \ref{fig:static_fault_tree_structure}.c) are used for convenience, to avoid extensive fault trees. The ``transfer in'' symbol is linked to its associated ``transfer out'', where the latter might develop elsewhere as part of the fault tree model.

%\begin{romanlist}

%    \item \textit{TRANSFER IN}: symbolized by a triangle with a line in its apex. 
    
%    \item \textit{TRANSFER OUT}: symbolized by a triangle with a line from the side.

%\end{romanlist}
Various FT extensions exist to model more complex dependability patterns, see \cite{vesely1981fault,dugan1992dynamic}. %for example \textit{transfer symbols}, used to make FT models manageable, further information of these gates can be found in \cite{vesely1981fault}. 

%\subsubsection{Extension of FTs}
% Different Types of Fault Trees
%FTs can be extended to other types of architectures employing additional logic, e.g., \textit{Dynamic Fault Tree Analysis} \cite{boudali2009rigorous}, \textit{Repairable Fault Tree Analysis}, \textit{State/Event Fault Tree Analysis}, among others.

% -----------------------------------
% Structural definition of a FT:
\subsection{Fault Tree Analysis}
% Say something about decision making in realiability engineering. That is the whole purpose of FTA.
FTA enables two types of analyses. \textit{Qualitative} analysis is based on the FT structure and aims at finding the critical system components.  \textit{Minimal cut sets} are minimal combinations of component failures that lead to a system failure. Small cut sets point to system vulnerabilities.
For example, a minimal cut set in Fig. \ref{fig:static_fault_tree_structure}.(c) is given by \textit{Contamination tape fails} (B1), and \textit{Basic cause seal failure} (B2). 

{\em Quantitative analyses} aim at computing various dependability metrics,  such as system \textit{reliability} (i.e., probability that the system fails in a period of time); the  \textit{availability} (i.e., the percentage of time that the system remains operational); \textit{mean time to failure} (i.e., average time before the first failure).
These metrics require the leaves of the FT to be equipped with failure probabilities, either as probability density functions or constant probabilities. An extensive list of algorithms for analysis of FTs is provided in \cite{ruijters2015fault}.
% -----------------------------------
% Design and development of a Fault Tree
% Design requirements of a FT:
\subsection{Construction of FTs}

Traditionally, FTs are handcrafted by experts on a system of interest. Automatic algorithms for induction of FTs are IFT \citep{madden1970generation}, LIFT \citep{nauta2018lift}, based on Evolutionary Algorithms \citep{linard2019fault}, Bayesian Networks \citep{linard2019induction}, ILTA \citep{waghen2019interpretable}, and DDFT \citep{Feng2020DATADRIVENFT}. Two main associated challenges are (i) discovering the FT structure that efficiently and completely represents the system failure mechanisms for a given top event, and (ii) finding patterns in a dataset containing information about the system failure mechanisms.

%\cite{lee1985fault} presents a list of algorithms used to build FTs. %Some examples are DRAFT \cite{fussell1973formal}, CAT \cite{salem1980cat}, by Lapp \& Powers \cite{lapp1977computer}, by Taylor \cite{hollo1976algorithms}.

%In general, these algorithms seek to identify correlations, determine the types of gates, resulting in the coupling of the tree in its final form. Among the techniques used are failure transfer functions, decision tables, directed graphs and cause and consequence diagrams.

% When inducing FTs in an automatic manner, similar to Decision Trees (DT) (Section \ref{sec:DT}), some stopping criteria are: (i) minimum gate size, (ii) maximum tree depth.

% If the top event is too general, the analysis become unmanagable, if it is too specific, the analysis does not provide a sufficiently broad view of the problem \cite{vesely1981fault}

% -----------------------------------
% Practical examples

\subsection{FT example}\label{sec:ft_example}

%Fig. \ref{fig:static_fault_tree_structure}.c presents an example adopted from \cite{stamatelatos2002fault} on a Fault Tree for a contamination and common-cause failure (CCF) Risk-Based design.

Fig.~\ref{fig:static_fault_tree_structure}.(c) shows a fault tree
 modeling the failure of a sealing mechanism for a container, adapted from \cite{stamatelatos2002fault}.
 The sealing function fails either due to a common cause of seal failure occurs or if the seals fail independently. For the former, it is necessary that the contamination tape fails and a basic seal failure occurs. For the latter, it is necessary for the metal-to-metal seal, the fused plug, and at least two of the three compression seals to fail. 

As shown in Fig.~\ref{fig:static_fault_tree_structure}.(c), the top event is refined in three intermediate events and independent basic events. This example, for simplicity, only considers \textit{AND}, \textit{OR} and \textit{VOT} gates. Since there is a quantification of the failure probability of each basic event, a \textit{qualitative} analysis can be carried out. Given the failure probabilities of the basic events in Fig.~\ref{fig:static_fault_tree_structure}.(c) and assuming independence between all basic events, the failure probability $P(G_1)$ of the top event {\em Sealing function fails} reads \[
P(G_1) = P(G_2) + P(G_3) - P(G_2)P(G_3) \:,
\]
where we have, with $p=10^{-3}$ the failure probability of the events $B_4$ through $B_6$,
\begin{eqnarray*}
P(G_4) & = & 3 p^2 (1-p) + p^3 \approx 3 \times 10^{-6} \\
P(G_3) & = & P(B_3) P(G_4) P(B_7) \approx 3  \times 10^{-12} \\
P(G_2) & = & P(B_1) P(B_2) = 1  \times 10^{-6} \: ,
\end{eqnarray*}
and hence $P(G_1) \approx 1  \times 10^{-6}$.

% -----------------------
\subsection{Benefits and limitations}
Some important advantages of FTs are that they (i) are based on probability theory, (ii) enable computation of different metrics (e.g., reliability) to aid decision making, (iii) due to its interpretability, they serve as a communication tool across multiple disciplines, helping to align stakeholders, (iv) facilitate the interpretation of different failure mechanisms, helping to identify critical components. 
Limitations of FTs include (i) the hand-made way in which they are traditionally constructed makes them costly and time consuming, (ii) the assumption that basic events are independent is not always fulfilled, (iii) validation in FTs is carried out based on expert judgment, which makes it subjective and prone to human error, (iv) difficult to collect appropriate data, (v) AND/OR gates are not always expressive enough. Some of the above aspects were also pointed out by \cite{sarbayev2019risk}.
\section{Decision Trees}\label{sec:DT}

% --------------------------------------------
\subsection{Purpose and application}

Decision Trees (DT) are flowchart-like structures that model decisions and their possible consequences. DTs are used by decision makers, mainly due to its intuitive interpretation. Technically, DTs serve as a classifier presented as a hierarchical top-down graph that generates a set of decision rules \citep{thomas2020machine}. DTs can cope with numerical and nominal attributes, and are often used in classification and regression problems. DTs are applied in text classification, diagnosis of diseases, fraud detection, speech recognition, video analysis, among others \citep{rokach2008data}. 

\begin{figure}[!h]
\centering
\includegraphics[width=1\linewidth]{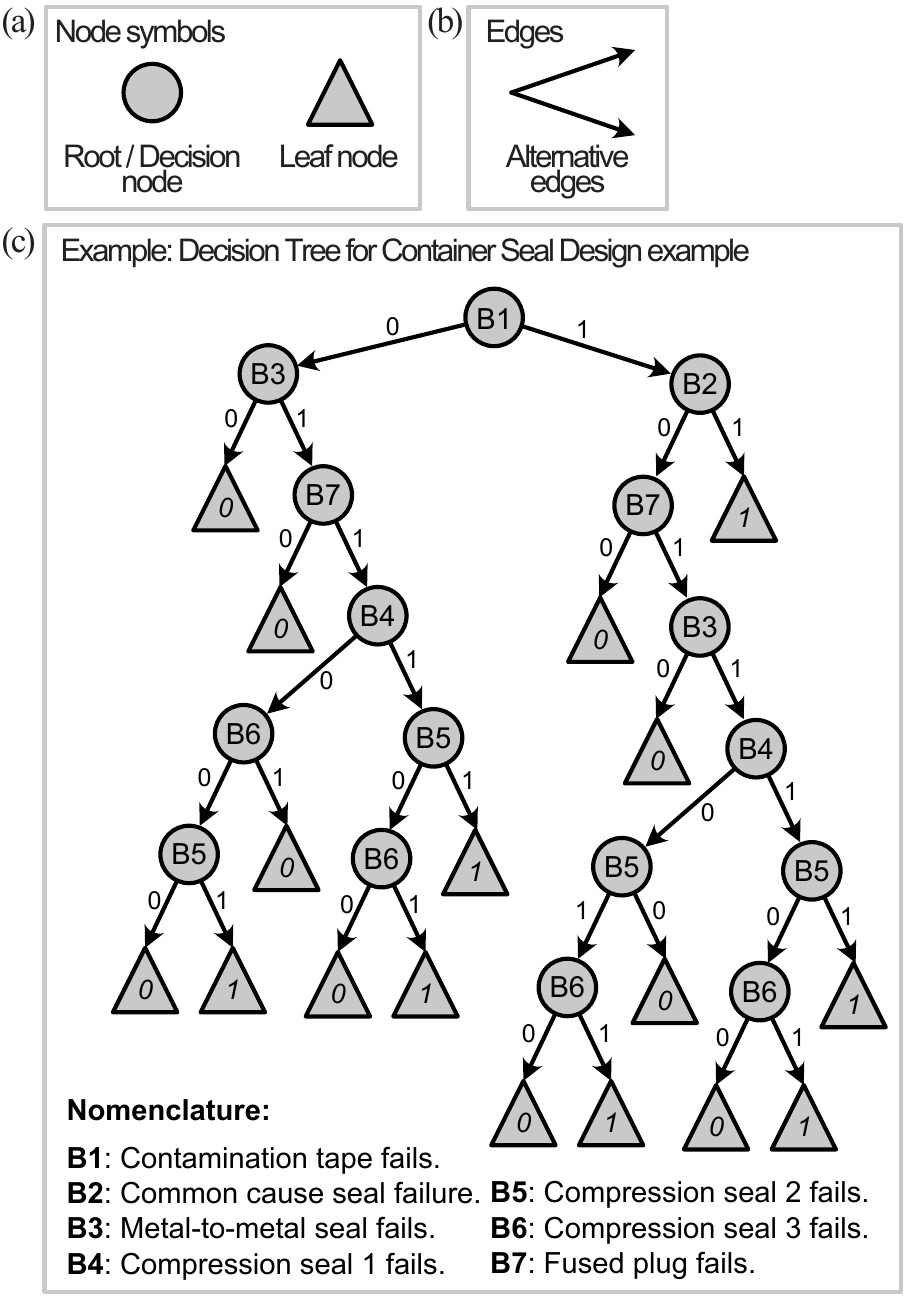}
\caption{(a) node symbols, (b) edges, (c) DT model corresponding to the FT from Fig.~\ref{fig:static_fault_tree_structure}.(c).}
\label{fig:decision_tree_elements}
\end{figure}

% --------------------------------------------
\subsection{Structural representation}

A DT is composed of nodes containing control statements based on attributes (or features). Among the common elements in a DT (Fig.~\ref{fig:decision_tree_elements}.(a-b)) are the (i) \textit{decision nodes} where the first on top is known as the \textit{root node}, (iii) incoming and outgoing \textit{edges}, and (iv) the \textit{leaf nodes} at the bottom of the structure representing the end point of a decision path. Each decision node is associated to an attribute, and the branches to a discrete category or to a range of values.

% --------------------------------------------
\subsection{Evaluation of DTs}

There are two main types of DTs: \textit{classification trees} and \textit{regression trees}. The first represents a function that maps all available samples (i.e., input data) into a predefined set of discrete categories, also known as \textit{labels}. The second attempts to predict a continuously-valued target attribute based on a set of input attributes. An example of how to evaluate the decision rules in a DT is presented in Section \ref{sec:pract_ex_DT}.

% --------------------------------------------
\subsection{Construction of DTs}

Induction algorithms, also called \textit{inducers}, build a model that generalizes the relationship between input attributes and target attributes based on a given training set of examples \citep{rokach2008data}. DT inducers look for the best feature upon which to perform split by means of \textit{splitting criteria}. Among the most commonly used are the \textit{Gini index}, and \textit{gain ratio}.
\textit{C4.5} \citep{quinlan1993c4} is a well-known DT induction algorithm. \cite{rokach2008data} provides a detailed list of other DTs inducers.

% --------------------------------------------
\subsection{DT example}\label{sec:pract_ex_DT}

To learn a DT that represents the same qualitative information as the FT in Fig~\ref{fig:static_fault_tree_structure}.(c), we proceeded as follows. First, we randomly generated 1000 data points, by drawing the basic events independently from a binomial distribution with probability of success equal to 0.5 and calculating the corresponding top event (0 for no failure and 1 for failure) by following the logical rules of the FT. The randomly drawn values for the basic events represent the input $x$ and the value of the top event the output $y$.

Then, a \textit{Binary Decision Tree} for classification was induced based on the above described dataset in a supervised fashion (i.e., mapping $x$ into $y$), using the function \textit{fitctree} in Matlab which uses the CART algorithm \citep{breiman1984classification}. The results are presented in Fig. \ref{fig:decision_tree_elements}.(c). The basic events in the FT (Fig. \ref{fig:static_fault_tree_structure}.(c)) become decision nodes in the DT. The leaf nodes in the DT are Boolean, representing the classes no failure and failure.

By interpreting the decision rules of the DT one could say that if the \textit{contamination tape} does not fail ($B_1=0$) and the \textit{metal-to-meal seal} does not fail ($B_3=0$) the system will not fail. Or if the \textit{contamination tape} fails ($B_1=1$) and the \textit{basic cause seal failure} occurs ($B_2=1$), the system fails. This DT encodes all cut sets of the FT, which can be obtained by concatenating all decision nodes with value 1 that lie on a path to a leaf node with value 1, yielding for example: \{$B_3$, $B_7$, $B_4$, $B_5$\}, \{$B_3$, $B_7$, $B_4$, $B_6$\}, etc. 

% --------------------------------------------
\subsection{Benefits and limitations of DT}

DTs have the following benefits (i) they are intuitive and easy to follow by technical and non-expert users, (ii) navigating through the branches associated to fault states enables the identification of useful logical rules (similar to the concept of cut sets in FTs), (iii) there are many algorithms for learning DTs from data that scale favorably with the number of data points. DTs have the following limitations (i) it is possible to have several identical sub-trees, which affects efficiency and computational performance. (ii) small variations in a splitting node located near the root of the tree can result in a completely new structure.

\section{Binary Decision Diagrams}\label{sec:BDD}

\subsection{Purpose and application}

Binary Decision Diagrams (BDDs) were originally introduced by \cite{lee1959representation} and are heavily used in model checking  \citep{bryant2018binary}, hardware design \& verification, protocol validation, and automated deduction \citep{rauzy1997exact}. BDDs also provide efficient algorithms to calculate the failure probabilities and minimal cut sets in a fault tree \citep{bryant1992symbolic,reay2002efficient}.

Syntactically, a BDD is a (connected and single-rooted) directed acyclic graph (see Fig. \ref{fig:binary_decision_diagram_elements}). It provides a very succinct representation and manipulations of Boolean expressions \cite{bryant1992symbolic}.

\subsection{Structural representation}
\begin{figure*}[!h]
\centering
\includegraphics[width=0.95\linewidth]{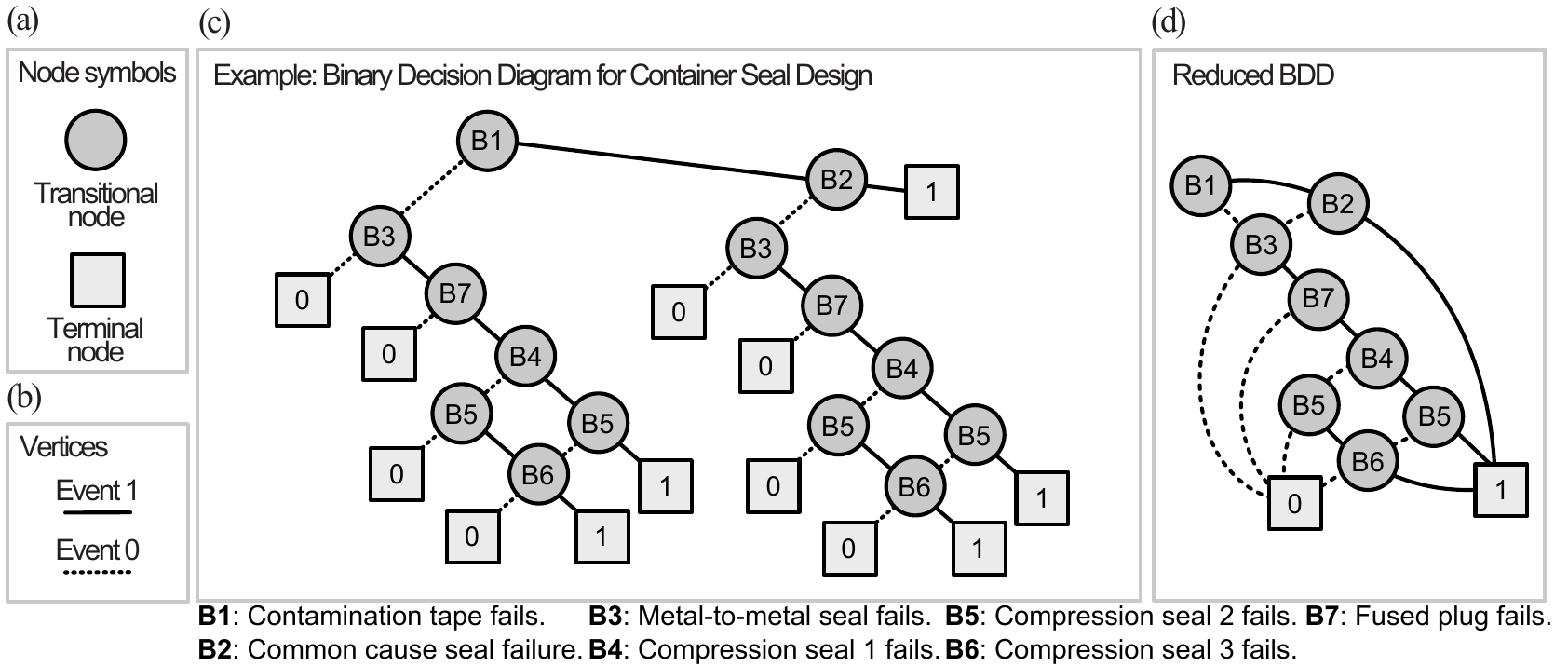}
\caption{(a) BDD node symbols, (b) BDD vertices, (c) BDD for FT from Fig. \ref{fig:static_fault_tree_structure}.(c), (d) Reduced BDD.}
\label{fig:binary_decision_diagram_elements}
\end{figure*}

BDDs consist of \textit{Transitional nodes}, or \textit{non-terminal vertices}. These are depicted by circles and contain binary control statements or \textit{function variables}. Terminal nodes, also called \textit{leaf nodes} or \textit{terminal vertices}, are represented by squares, and are labeled with either $1$: \textit{True} or, $0$: \textit{False}. Conventionally, the vertices (Fig. \ref{fig:binary_decision_diagram_elements}.(b)) are depicted as a solid line if the output of the transitional node is $1$, and dashed lines if the output is $0$.

\subsection{Evaluation of BDDs}\label{sec:evaluations_BDDs}

A reason for the popularity of BDDs is their efficient handling of many operations on Boolean functions, including 
 negation, conjunction  and disjunction: 
 Given BDD representations of $F$ and $G$ over the same variable ordering, efficient algorithms exist to compute a compact BDD for $\neg F$, $F\land G$ and $F\lor G$ \citep{bryant1986graph}.

Given a BDD over a set of variables $V$, one easily evaluates this BDD over an assignment of $V$: 
One starts from the root, and for each variable $v\in V$, one takes the left branch / solid line if $v=1$, and right branch / dashed line if $v=0$. Eventually, one reaches a leaf yielding the output for this assignment.

\subsection{Construction of BDDs}\label{sec:construction_of_bdds}

Given a Boolean function $F$,
the BDDs representation for $F$ can be constructed recursively applying the \textit{Shannon expansion formula} \citep{akers1978binary}
\[
F(x_1,x_2,...) = x_1F(1,x_2,...) \vee (1-x_1)F(0,x_2,...) \:
\]
Here $F$ is a Boolean function; and $x_i$ are binary values. In addition, compact representations of BDDs (known as reduced BDDs) for a given order of variables can be obtained using \textit{collapsing operations} by eliminating/merging nodes and sub-diagrams \citep{friedman1987finding}.

\subsection{BDD example}\label{sec:practical_ex_BDD}

By applying the \textit{Shannon expansion formula} (Section \ref{sec:construction_of_bdds}), we convert the FT in Fig. \ref{fig:static_fault_tree_structure}.(c) into a BDD (Fig. \ref{fig:binary_decision_diagram_elements}.(c)), with the associated Boolean function:
\begin{equation*}
\begin{aligned}
F_1(B_1, B_2,..., B_7)  = {} & F_2(B_1,B_2) + F_3(B_3,...,B_7) \\
        F_2(B_1,B_2)   = {} & B_1 \cdot B_2  \\      
    F_3(B_3,...,B_7)   = {} & B_3 \cdot B_7 \cdot (B_4 \cdot B_5 + \\
                        & B_4 \cdot B_6 + B_5 \cdot B_6) 
\end{aligned}
\end{equation*}
Here $F_2$ and $F_3$ are Boolean functions that model respectively the intermediate events in Fig.~\ref{fig:static_fault_tree_structure}.(c). One can see great similarity between this BDD and the DT in Fig.~\ref{fig:decision_tree_elements}.(c), except that a BDD allows shared nodes and the DT cannot.

Through collapsing operations, like merging the nodes B3 and all the terminal nodes, the \textit{Reduced BDD} in Fig. \ref{fig:binary_decision_diagram_elements}.(d) is obtained. The latter, easily identify the minimum cut sets, i.e., all 1-edge vertices that connect to the terminal node of value 1: $\{B_1, B_2\}$, $\{B_3,B_4,B_5,B_7\}$, $\{B_3,B_5,B_6,B_7\}$, and $\{B_3,B_4,B_6,B_7\}$.

\subsection{Benefits and limitations}
Among the advantages offered by BDDs are that (i) they enable a representation of any Boolean function; (ii) there is a wide variety of algorithms that allow performing operations of Boolean functions. One limitation is that finding the order of variables that minimizes the resulting BDD is an NP-hard problem \citep{bollig1996improving}.

% -------------------- %
\section{Conversion methods}\label{sec:conversion_methods}
\begin{figure}[!h]
\centering
\includegraphics[width=0.89\linewidth]{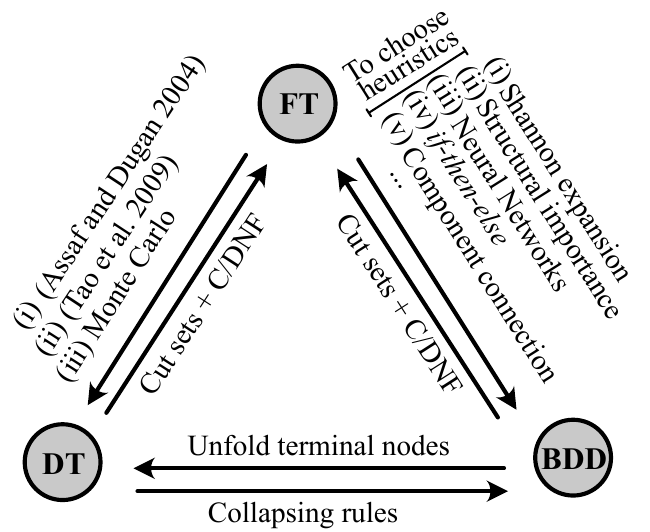}
\caption{Conversion methods among FT, DT, and BDD.}\label{fig:transformation_algorithms}
\end{figure}

Conversion methods are mathematical transformations that convert one formalism into another, while preserving relevant properties. So we look for those that allow transitions between FTs, DTs and BDDs (see Fig. \ref{fig:transformation_algorithms}).

In the transformation \textbf{FT $\to$ BDD}, the basic events in the FT become transitional nodes in the BDD, and the top event in the FT is represented in the terminal nodes of the BDD. We identified a few methods where the main idea is to recursively apply the \textbf{(i)} \textit{Shannon expansion} until all basic events are converted into BDD nodes \citep{akers1978binary}. Since the variable ordering has a crucial effect on the size of a BDD, a number of heuristic approaches have been developed to find a good order. Examples include \textbf{(ii)} \textit{structural importance} of each basic event \citep{bartlett2001ordering}, \textbf{(iii)} a \textit{neural network} approach to choose the best heuristic from a set of alternatives \citep{bartlett2002choosing}, \textbf{(iv)} the \textit{if-then-else} method that focuses on the gates in the fault tree \citep{remenyte2008enhanced}, and \textbf{(v)} the \textit{component connection method}, where gates with only basic events as inputs are considered \citep{remenyte2008enhanced}.

In the transformation \textbf{BDD $\to$ FT}, the terminal nodes of the BDD become the top event of the FT, and the transitional nodes of the BDD the basic events of the FT. We did not find specific techniques in the literature that accomplish this transition. Not every BDD can be transformed into a FT, since FTs do not support a NOT function. One approach could be to obtain the cut sets from the BDD and then reconstruct the associated FT e.g., by means of the (con/dis)junctive normal form (C/DNF). Since the resulting FT may not be optimal (e.g., redundant elements), minimization rules should be applied on the FT \citep{junges2017fault}.

In the transformation \textbf{FT $\to$ DT} the basic events in the FT become decision nodes in the DT, and the top event of the FT the leaf nodes in the DT. A few methods have been developed to accomplish this task. \textbf{(i)} \cite{assaf2004diagnostic} translate a dynamic fault tree (DFT) to a Markov chain model, and then to a Diagnostic Decision Tree (DDT). \textbf{(ii)} \cite{tao2009notice} compute from the fault tree the minimal cut sets, top event probabilities and the \textit{Vesely-Fusell} measure, and with this information build the DDT. One may also apply \textbf{(iii)} the method that we applied in Section \ref{sec:pract_ex_DT} to the running example: randomly generate a binary data set using the FT (i.e., Monte Carlo method), and then learn the corresponding DT.

In the transformation \textbf{DT $\to$ FT} the decision nodes in the DT become basic events in the FT, and the leaf nodes in the DT the top events in the FT. We did not find any publication to accomplish this transition. Nevertheless, by using all decision rules that connect to a leaf node of value 1, one can get cut sets and build the FT as suggested in the transition BDD $\to$ FT.

The transformation \textbf{DT $\leftrightarrow$ BDD} is trivial. For \textbf{DT $\to$ BDD} \textit{collapsing operations} are needed, which aim at eliminating redundant nodes, redirecting the edges, and applying isomorphism rules to eliminate transitional nodes with two terminal nodes of same value \citep{ZhengHao:online}. Vice versa, \textbf{BDD $\to$ DT} can be achieved by \textit{unfolding} the terminal nodes in the BDD until each node has a single parent node (except for the root node).
% -------------------- %
\section{Discussion}\label{sec:discussion}
Section \ref{sec:conversion_methods} provides different algorithms that enable the transition FT $\to$ BDD. By representing FTs as BDDs, standard techniques can be applied for qualitative and quantitative analysis. The transition FT $\to$ DT has been proposed to obtain Diagnostic Decision Trees (DDTs), which are arguably easier to interpret by non-expert users. We did not find any official publication that addresses the transition BDD $\to$ FT or DT $\to$ FT, we suspect this is because it is not clear what the added value of the transition BDD $\to$ FT is, and that making the transition DT $\to$ FT needs further exploration. However, a way to carry out such transformation is by first obtaining the cut sets from either the BDD or the DT, and then build the FT in its (con/dis)junctive normal form. Since, the resulting FT may be sub-obtimal, it is necessary to apply minimization rules. The transition in both sides of binary DT $\leftrightarrow$ BDD is straightforward since a binary DT is a special case of a BDD, where the former is less efficient than a reduced ordered BDD.

In Table \ref{tb:comparision_between_models} we summarized the aspects to compare between models. Although all three models were born in graph theory, they serve very different purposes and application domains. Consequently, they adopt different structural requirements and analysis methods. Regarding the construction of these models, DTs stands out by the availability of efficient induction algorithms, FTs on the other hand are mainly ``handmade''. An ordered BDD and a binary DT have the same structure, but since BDDs allow node-sharing, they can be more compact than binary DTs. All three models carry all sorts of benefits and limitations. The limitations can be addressed via extensions or conversion methods.
% -------------------- %
\section{Conclusions}\label{sec:conclusions}
We compared three well-known graph models, fault trees (FT), decision trees (DT) and binary decision diagrams (BDD). We used a running example to exemplify the properties offered by each model, and discuss conversion methods to transition from one model to the formulation of another. We observed that the transformation FT $\to$ BDD is well investigated, the transformations DT $\to$ FT or BDD $\to$ FT are not investigated and we briefly discussed how to carry them out, and that the transformation DT $\leftrightarrow$ BDD is trivial. We conclude that, given their different purposes and application domains, these models adopt different structural representations and analysis methodologies that entail a variety of benefits and limitations. These limitations may be addressed via conversion methods or extensions. Specific remarks are that BDDs can be considered as a compact representation of binary DTs, since the former allow sub-node sharing, which makes BDDs more efficient at representing logical rules than binary DTs. It is possible to obtain cut sets from BDDs and DTs and construct a FT using the (con/dis)junctive normal form, although this may result in a sub-optimal FT structure.

% -------------------- %

\medskip
\noindent
{\bf Acknowledgement.}
This research has been partially funded by NWO under the grant PrimaVera (https://primavera-project.com) number NWA.1160.18.238.

% -------------------- %

\bibliography{sample}
\bibliographystyle{chicago}

% -------------------- %
\end{document}